\documentclass[twocolumn]{aastex631}
\usepackage{chemformula}

\received{March 31, 2023}
\revised{July 21, 2023}
\accepted{August 4, 2023}

\submitjournal{PSJ}

\begin{document}

\title{The Spatial Distribution of the Unidentified 2.07 \textmu m Absorption Feature on Europa and Implications for its Origin}

% \correspondingauthor{M. Ryleigh Davis}
% \email{rdavis@caltech.edu}

\author[0000-0002-7451-4704]{M. Ryleigh Davis}
\affiliation{Division of Geological and Planetary Sciences, California Institute of Technology, Pasadena, CA 91125, USA}

\author[0000-0002-8255-0545]{Michael E. Brown}
\affiliation{Division of Geological and Planetary Sciences, California Institute of Technology, Pasadena, CA 91125, USA}

\author[0000-0002-0767-8901]{Samantha K. Trumbo}
\affiliation{Cornell Center for Astrophysics and Planetary Science, Cornell University, Ithaca, NY 14853, USA}

\begin{abstract}
A weak absorption feature at 2.07 \textmu m on Europa's trailing hemisphere has been suggested to arise from radiolytic processing of an endogenic salt, possibly sourced from the interior ocean. However, if the genesis of this feature requires endogenic material to be present, one might expect to find a correlation between its spatial distribution and the recently disrupted chaos terrains. Using archived near-infrared observations from Very Large Telescope/SINFONI with a $\sim$1 nm spectral resolution and a linear spatial resolution $\sim$130 km, we examine the spatial distribution of this feature in an effort to explore this endogenic formation hypothesis. We find that while the presence of the 2.07 \textmu m feature is strongly associated with the irradiation pattern on Europa's trailing hemisphere, there is no apparent association between the presence or depth of the absorption feature and Europa's large-scale chaos terrain. This spatial distribution suggests that the formation pathway of the 2.07 \textmu m feature on Europa is independent of any endogenous salts within the recent geology. Instead, we propose that the source of this feature may simply be a product of the radiolytic sulfur cycle or arise from some unidentified parallel irradiation process. Notably, the 2.07 \textmu m absorption band is absent from the Pwyll crater ejecta blanket, suggesting that radiolytic processing has not had enough time to form the species responsible and placing a lower limit on the irradiation timescale. We are unable to find a plausible spectral match to the 2.07 \textmu m feature within the available laboratory data.
\end{abstract}

%% New Keyword definitions at: http://astrothesaurus.org

\section{Introduction} \label{sec:intro}

 Beneath its icy shell, Jupiter's moon Europa hosts a deep global salty ocean that is likely in direct contact with the silicate mantle. When combined with a possible supply of radiolytically produced oxidants from the surface, this water-rock interaction makes Europa a prime target for astrobiological consideration \citep[e.g.][]{anderson1998_EuropaDifferentiatedInternal, kivelson2000_GalileoMagnetometerMeasurements, chyba2000_EnergyMicrobialLife, hand2009_AstrobiologyPotentialLife}. Our current best window into Europa's largely unconstrained ocean chemistry is through the surface composition. Specifically, the recently geologically active chaos terrains, bands, and ridges show clear spectroscopic evidence for the presence of salt hydrates that are likely sourced from the subsurface ocean \citep[e.g.][]{mccord1999_HydratedSaltMinerals, johnson2002_EnergyDistribution, leblanc2002_EuropaSodiumAtmosphere, dalton2005_SpectralComparisonHeavily, shirley2010_EuropaRidgedPlains, trumbo2019_SodiumChlorideSurface, trumbo2020_EndogenicExogenicContributions}. These large scale chaos terrains, characterized by polygonal iceberg like blocks of ice within a matrix \citep{carr1998_EvidenceSubsurfaceOceana}, are interpreted to be areas of focused heat flow and possibly local melting which may have exposed oceanic material \citep[e.g][]{greenberg1999_ChaosEuropa, collins2009chaotic}. However, exogenic sulfur ions from Io's volcanoes are continuously deposited onto Europa's trailing hemisphere, and bombardment by magnetospheric ions, protons, and electrons \citep{pospieszalska1989_MagnetosphericIonBombardment, cooper2001_EnergeticIonElectron, paranicas2001_ElectronBombardmentEuropa, paranicas2009_EuropaRadiationEnvironment} drives a radiolytic sulfur cycle. This cycle chemically alters the surface composition and creates a ``bull’s-eye" pattern of hydrated sulfuric acid centered at the trailing hemisphere apex (270$^{\circ}$W, 0$^{\circ}$N) \citep{carlson1999_SulfuricAcidEuropa, carlson2002_SulfuricAcidProductiona, carlson2005_DistributionHydrateEuropaa}. Understanding the source and nature of exogenic material and radiolytic processing on Europa's surface and disentangling the original state of the endogenic material is therefore critical for interpreting Europa's surface composition and inferring the properties of its underlying ocean.

Europa's surface composition includes crystalline and amorphous water ice \citep{hansen2004amorphous}, radiolytically produced hydrated sulfuric acid on the trailing hemisphere \citep{carlson1999_SulfuricAcidEuropa, carlson2005_DistributionHydrateEuropaa}, and an additional hydrated salt component that is largely constrained to the recently geologically active chaos terrains, bands, and ridges \citep{mccord1999_HydratedSaltMinerals, dalton2005_SpectralComparisonHeavily, daltoniii2012_EuropaIcyBright}. The distribution of this additional salt hydrate is consistent with Europa's recent geologic features containing endogenic material sourced from the subsurface. Irradiation produced \ch{O2} and hydrogen peroxide (\ch{H2O2}) \citep{johnson2003_ProductionOxidantsEuropa, loeffler2006_SynthesisHydrogenPeroxide, trumbo2019_H2O2ChaosTerrain} as well as \ch{CO2} \citep{mccord1998_NonwatericeConstituentsSurface, hansen2008_WidespreadCO2Other, carlson2009_EuropaSurfaceComp} have also been confirmed on Europa's surface. Recent observations with HST/STIS revealed the presence of absorption features at 230 and 450 nm, spatially constrained to the leading hemisphere chaos terrains \citep{trumbo2019_SodiumChlorideSurface, trumbo2022_NewUVSpectral}, which closely match color center features produced in laboratory data of irradiated sodium chloride (\ch{NaCl}) \citep{hand2015_EuropaSurfaceColor, denman2022_InfluenceTemperaturePhotobleachinga, brown2022_MidUVSpectrumIrradiateda}. Together, these results provide strong evidence for the presence of endogenous \ch{NaCl} in Europa's leading hemisphere chaos. 

Additional endogenic salt species, which are spectroscopically distinct from the hydrated sulfuric acid, may also exist within Europa's chaos terrains \citep[e.g.][]{fischer2015_SpatiallyResolvedSpectroscopy, trumbo2020_EndogenicExogenicContributions}, however their precise nature is widely debated. Some analyses of Galileo/NIMS spectra have suggested that Europa's endogenous hydrate may be dominated by sulfate salts \citep[e.g.][]{mccord1998_NonwatericeConstituentsSurface, dalton2005_SpectralComparisonHeavily, daltoniii2012_EuropaIcyBright}, while more recent studies with higher spectral resolution data sets suggest a chloride dominated composition \citep{brown2013_SALTSRADIATIONPRODUCTS, ligier2016_VLTSINFONIOBSERVATIONS}. Further complicating the picture, the hydrated material of the leading hemisphere chaos is spectroscopically distinct from the hydrated material in the trailing hemisphere chaos, presumably due to the significant exogenous alteration which occurs on the trailing hemisphere \citep{fischer2015_SpatiallyResolvedSpectroscopy, fischer2016_SPATIALLYRESOLVEDSPECTROSCOPY, trumbo2020_EndogenicExogenicContributions}. \citet{trumbo2020_EndogenicExogenicContributions} demonstrated that a 360 nm absorption feature and 700 nm slope change strongly correlate to the trailing hemisphere chaos terrains and reflect a combination of endogenous and exogenous processes. They suggest that while relatively unaltered endogenic material may persist in the comparatively sheltered leading hemisphere chaos terrains, it may become progressively more chemically altered towards the trailing hemisphere apex due to radiolytic processing.

A weak absorption feature at 2.07 \textmu m is also present on Europa’s trailing hemisphere. It was first noted in the high resolution Keck/OSIRIS observations of \citet{brown2013_SALTSRADIATIONPRODUCTS}, who suggested epsomite (\ch{MgSO4 . 7 H2O}) or magnesium sulfate brine as a plausible spectral match. With no sign of a sulfate absorption on Europa’s leading hemisphere, they suggested that magnesium chloride (\ch{MgCl2}) sourced from Europa’s interior may be radiolytically converted into epsomite, a hydrated magnesium sulfate (\ch{MgSO4}), when in the presence of Iogenic sulfur ions deposited onto the trailing hemisphere. While it is possible that this conceptual model of the alteration of chloride-rich endogenic salts into sulfates via sulfur radiolysis could be consistent with the visible-wavelength data of the trailing hemisphere \citep{hibbitts2019_ColorCentersSulfates, trumbo2020_EndogenicExogenicContributions}, more laboratory data is needed to test this hypothesis. \citet{ligier2016_VLTSINFONIOBSERVATIONS}, on the other hand, used VLT/SINFONI observations to show that the 2.07 \textmu m absorption feature can also be reproduced by certain combinations of \ch{MgCl2} and perchlorate (\ch{Mg(ClO4)2}) salts, despite neither one having a distinct absorption on their own. The authors suggested that endogenic magnesium-chloride may be radiolytically converted into magnesium chlorate and perchlorate, rather than sulfate, on Europa's trailing hemisphere.

Both proposed origins for the 2.07 \textmu m absorption feature (magnesium-bearing sulfates versus chlorine salts) share one important characteristic. They require a contribution of endogenic material, presumably tied to the ocean chemistry, which is chemically altered into the species responsible for the absorption by the trailing hemisphere irradiation environment. In both of these scenarios we would expect to find a spatial correlation between the presence and depth of the 2.07 um feature and the large-scale geologic units or chaos terrains, where we would expect geologic resurfacing to have emplaced endogenic, salty material. The trailing hemisphere chaos terrains are spectroscopically distinct from their leading hemisphere counterparts \citep{trumbo2020_EndogenicExogenicContributions}, presumably due to irradiation induced chemical alteration of this endogenic material. Indeed, the best fit spectral models from \citet{ligier2016_VLTSINFONIOBSERVATIONS} show an enhancement of \ch{MgCl2} in the chaos terrains on both hemispheres and magnesium perchlorate confined mainly to the trailing hemisphere chaos, consistent with the 2.07 \textmu m feature arising from a mixture of endogenic chloride and its irradiation product. However, their models do not map the 2.07 \textmu m absorption feature directly, but rather rely on linear mixture modeling of the largely featureless near infrared continuum, and degeneracies amongst the various salt hydrates make positive identification difficult. A more recent study by \citet{king2022_CompositionalMappingEuropa} combined VLT/SPHERE and Galileo/NIMS observations in an attempt to model Europa's surface composition across the near-infrared. Their best fit models include sodium chloride and show distributions for both magnesium sulfate and magnesium bearing chlorine species, which are broadly consistent with both hypotheses. However, they point out that the uncertainties in the model abundances and degenerate fit results for various compositional mixtures mean confident detection of any individual salt species is not feasible with the existing data sets. 

Previous observations indicate that irradiated NaCl is likely present and spatially concentrated in the leading hemisphere chaos terrains \citep{trumbo2022_NewUVSpectral}, and confirmation of endogenic \ch{MgCl2} or its radiolytic products could provide crucial information for constraining Europa's ocean and surface chemistry. While atomic sodium and potassium, as well as chlorine ions, likely sputtered from the surface, have been detected in Europa's atmosphere or a nearby pickup cloud \citep{brown1996_DiscoveryExtendedSodium, brown2001_PotassiumEuropaAtmosphere, volwerk2001_WaveActivityEuropa}, attempts to measure magnesium in Europa's atmosphere were unsuccessful. \citet{horst2013_SEARCHMAGNESIUMEUROPA} provide an upper limit on the atmospheric \ch{Na}/\ch{Mg} ratio of at least a factor of two less than meteoritic or cosmic abundances, and the presence of magnesium bearing salts on Europa's surface remains unproven. Thus, the detection of \ch{MgCl2} along with \ch{NaCl} on Europa's surface may have important implications for the relative balance of various cation and anion species within Europa's ocean. In the relatively simple freezing models of \citet{johnson2019_InsightsEuropaOcean}, for example, the presence of \ch{MgCl2} could suggest a sulfate-poor, sodium, magnesium and chloride rich ocean with a low pH due the presence of magnesium ions. 
 
In this paper, we use high spatial and moderate spectral resolution archived VLT/SINFONI data to explore the spatial distribution of the 2.07 \textmu m feature and determine whether its presence is associated with the large-scale chaos terrains, which could confirm its suspected relationship to endogenic salts and possibly Europa's ocean chemistry. With two possible origins of Europa's 2.07 \textmu m absorption feature arising from radiolytic processing of endogenic \ch{MgCl2} and no clear resolution for the genesis of the feature from spectral modeling, we turn to the spatial distribution of the absorption across Europa's trailing hemisphere in an attempt to gain insight.

\begin{figure}[t!]
\plotone{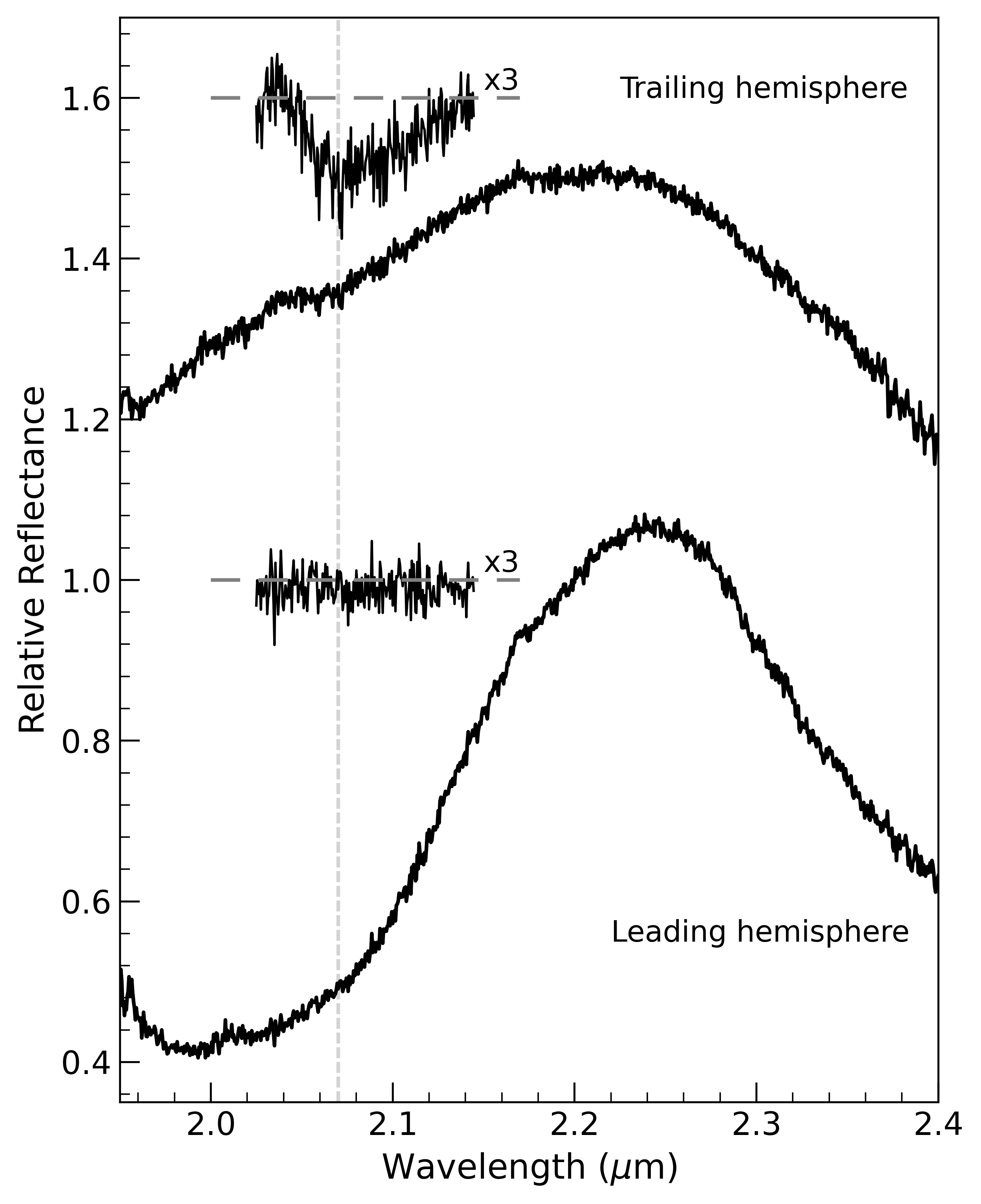}
\caption{Example VLT/SINFONI spectra of Europa's leading and trailing hemispheres showing the 2.07 \textmu m absorption feature, which is only present on Europa's more heavily irradiated trailing hemisphere. The data are normalized to 1.0 at 2.2 \textmu m and the trailing hemisphere spectrum is offset by +0.5 for clarity. The vertical dotted line marks the location of the 2.07 \textmu m band center and the continuum-removed absorption (multiplied by a factor of 3) is included above each spectrum. 
\label{fig:specs}}
\end{figure}

\section{Observations \& Data Reduction} \label{sec:data}

We use archived spatially resolved, moderate-resolution, ground-based near-infrared spectra from VLT/SINFONI acquired with adaptive optics during ESO program 088.C-0833(A) at Paranal Observatory and originally published in \citet{ligier2016_VLTSINFONIOBSERVATIONS}. The data were obtained by \citet{ligier2016_VLTSINFONIOBSERVATIONS} over five nights between October 2011 and January 2012 in the H+K band setting, covering a wavelength range of 1.452 - 2.447 \textmu m at a spectral resolution of $\sim$1 nm ($R \sim \frac{\lambda}{\Delta \lambda} = 1500$). Europa's angular diameter on-sky ranged from 0.897" to 1.083", so to cover the full disk with the 0.8" $\times$ 0.8" field of view, integrations were taken in five offset frames and co-added into a single mosaic across the full disk. The on-sky spatial sampling of the SINFONI instrument is 12.5 $\times$ 25 mas, resulting in spatial pixels of $\sim$ 35 $\times$ 70 km at the center of Europa's disk. We note that the spatial scale of the SINFONI pixels over-samples the diffraction-limited spatial resolution of the VLT at the wavelengths of interest, which is $\sim$ 130 km at the center of Europa. More detailed information on these observations can be found in Table 1 in \citet{ligier2016_VLTSINFONIOBSERVATIONS}.

We downloaded the raw data and associated calibration files from the ESO Science Archive Facility and followed a similar reduction routine as \citet{ligier2016_VLTSINFONIOBSERVATIONS}. As a first step in the data reduction process, we used the ESO SINFONI data reduction pipeline v3.3.3 with EsoRex \citep{modigliani2007sinfoni}. The pipeline identified bad, non-linear and hot pixels, performed standard dark and flat lamp corrections, corrected for optical distortions in the image slicer slitlets, determined a calibrated wavelength solution, and computed the 2D to 3D mapping to create a 3D data cube with two spatial dimensions and one spectral dimension for each observation. We interrupted the pipeline before the five offset frames were combined into a single mosaic to remove correlated regions of bad-pixel patches, which were not identified by the pipeline, and which were also noted by \citet{ligier2016_VLTSINFONIOBSERVATIONS}. We identified these bad pixel regions using the spectral angle mapping criterion (SAM) \citep{kruse1993spectral}, which describes the degree of similarity between two spectra. We masked out pixels with a SAM angle that differed by more than $\sim$10\% from the median SAM angle between all individual spectra and the combined median in a 10$\times$10 pixel bounding box centered on the bad pixel. These bad pixel regions, which makeup $\sim$7\% of the total detector pixels, were excluded from the final co-added data cube created after running the remainder of the ESO pipeline. While \citet{ligier2016_VLTSINFONIOBSERVATIONS} do not indicate what method they used to identify these remaining bad pixel patches, we find that our identified bad pixel patches closely match the pixels which appear to be masked in Figure 3 of \citet{ligier2016_VLTSINFONIOBSERVATIONS}.

Next, we used the standard star observations of the solar-type star HD18681, which were taken immediately before or after the Europa observations on each night, to correct for the instrument response, remove telluric absorption features, and derive a relative reflectance calibration. Due to the range of air masses over which the five Europa cubes and the calibration star were observed each night, an on-sky angular separation between Europa and HD18681 of $\sim$17$^\circ$, and possibly changing atmospheric conditions, we found a simple division by the telluric standard was not sufficient. The quality of the telluric correction is particularly important for our analysis of the 2.07 \textmu m absorption, which is just beyond telluric \ch{CO2} and \ch{O2} absorptions between 2.0 and 2.05 \textmu m. We therefore followed a similar telluric correction procedure as described in \citet{brown2013_SALTSRADIATIONPRODUCTS} and continuum divided the observed telluric spectrum by a best-fit polynomial to construct an atmospheric transmission spectrum. We then empirically determined an exponential scaling factor which adjusts the depths of the telluric absorptions in the transmission spectrum to most closely match the telluric signature in each of the Europa observations, paying particular attention to the quality of the telluric removal of the \ch{CO2} and \ch{O2} features between 2.0 and 2.05 \textmu m. Each observation consists of hundreds of spectra, which allowed us to determine the optimal scaling factor to very high precision. The original stellar continuum fit was multiplied by this best-fit atmospheric transmission spectrum to create a new standard star spectrum with an appropriately scaled telluric signature, which was then divided from the Europa observations. As described in \citet{ligier2016_VLTSINFONIOBSERVATIONS}, we used a Lambertian surface to calculate the geometrically corrected reflectance spectrum of each pixel and corrected for small offsets in the relative reflectance of overlapping regions on Europa's surface obtained during different nights by scaling the band-integrated flux of each observation cube so that the integrated reflectance matched in these overlapping regions. The final data product is a series of data cubes, one for each hemispherical view on Europa (centered at 30$^\circ$W, 55$^\circ$W, 130$^\circ$W, 225$^\circ$W, and 315$^\circ$W) with the corresponding relative reflectance spectrum for each pixel \footnote{The full, reduced dataset is permanently archived at https://doi.org/10.22002/agga7-v3393}. A characteristic leading and trailing hemisphere spectrum, at (90$^\circ$W, 0$^\circ$N) and (270$^\circ$W, 0$^\circ$N) respectively, are shown in Figure \ref{fig:specs}, with the 2.07 \textmu m absorption feature present only in the trailing hemisphere spectrum.

\begin{figure*}[ht!]\epsscale{0.88}
\plotone{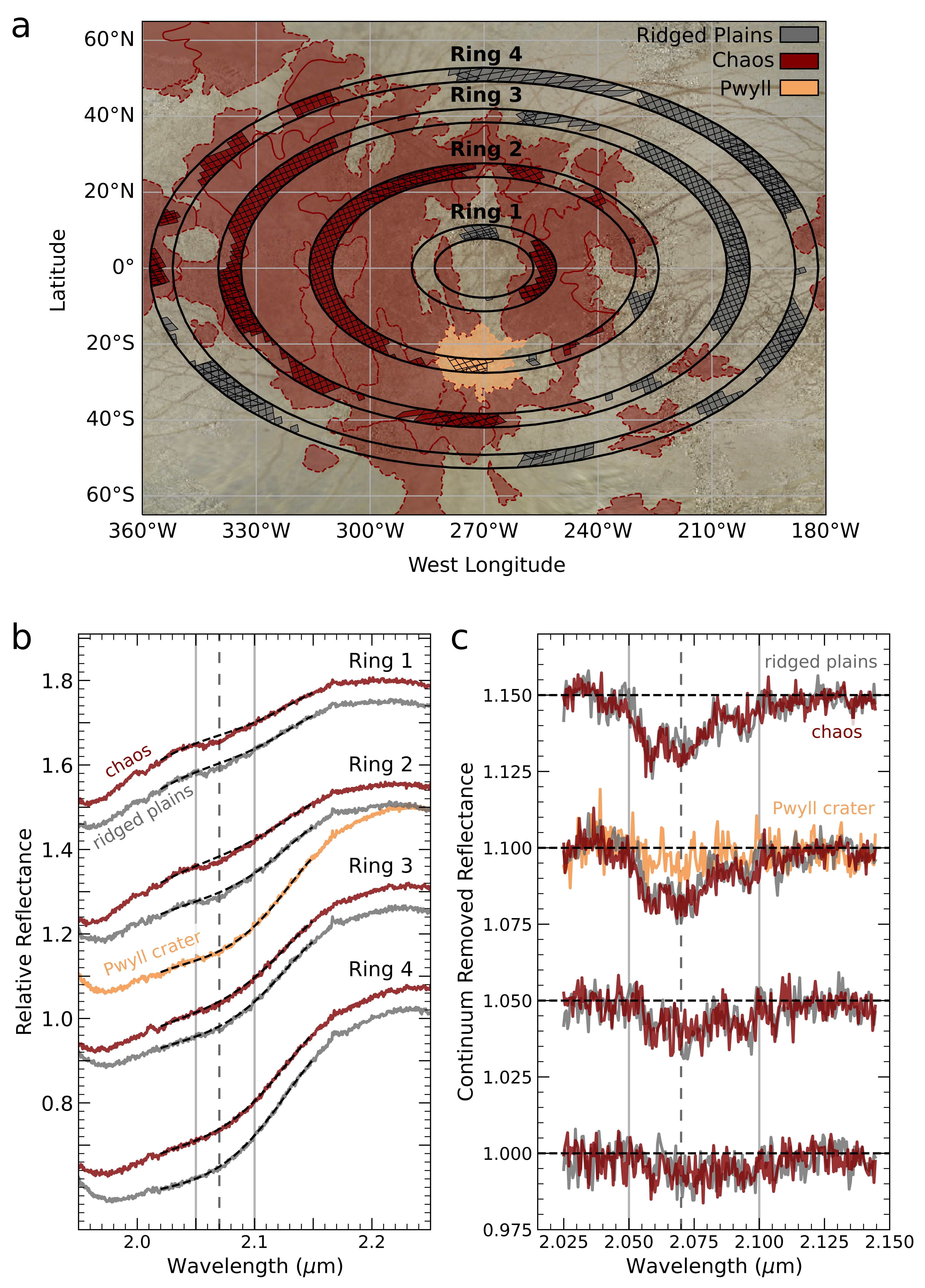}
\caption{(a) A map of Europa's trailing hemisphere, showing a selection of four of the 29 considered ``rings of constant radiolysis" outlined in black, which are used to determine whether the weak 2.07 \textmu m absorption feature is associated with Europa's large-scale geologic units. The chaos terrains are shown in red and Pwyll crater and its ejecta are orange. For each ring, projected detector pixels are shown and their assigned geologic unit indicated by color, where chaos terrains are dark red, Pwyll crater and its ejecta are orange, and the ridged plains are gray. (Europa map image credit: NASA/JPL/ Björn Jónsson).
(b) Summed spectra, by geologic unit, for all of the pixels within each example ring. Spectra are normalized to 1.0 at 2.12 \textmu m and offset for clarity. The best-fit polynomial continuum is indicated by a black dashed line.
(c) Continuum removed absorption feature, by geologic unit, for the four selected example constant radiolysis rings. The depth of the feature decreases noticeably as we move from Ring 1 outwards, however there is no apparent difference in the depth of the 2.07 \textmu m feature between the chaos terrains (red) and ridged plains (gray) for a given ring. This suggests that the absorption feature is unaffected by the compositional difference between the chaos terrains and ridged plains, and it is therefore unlikely to be related to any of the proposed irradiation products of endogenic material. Instead, the 2.07 \textmu m feature appears to arise from the radiolytic processing of water ice with exogenic material, possibly related to the sulfur radiolytic cycle. One notable exception is that the 2.07 \textmu m absorption feature is entirely absent from Pwyll crater and its ejecta blanket.}\label{fig:ring}
\end{figure*}

\begin{figure}[t!]\epsscale{1}
\plotone{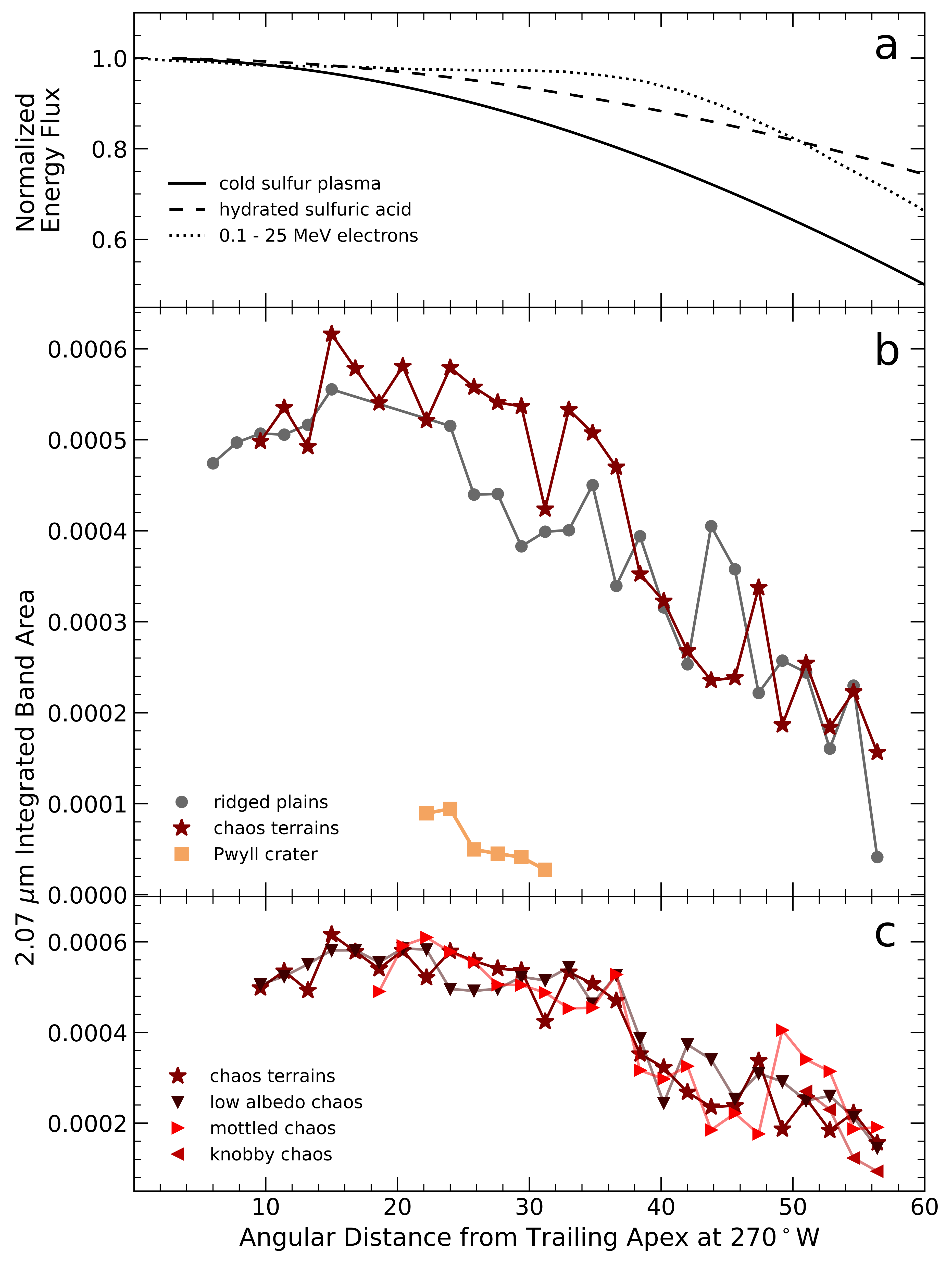}
\caption{(a) Expected trailing hemisphere irradiation patterns as a function of the angular distance from the trailing hemisphere apex for cold sulfur ions (solid) \citep{cassidy2013_MagnetosphericIonSputtering}, radiolytically produced hydrated sulfuric acid (dashed) \citep[e.g.][]{carlson2002_SulfuricAcidProductiona, ligier2016_VLTSINFONIOBSERVATIONS}, and 0.1 - 25 MeV electrons (dotted) \citep{paranicas2009_EuropaRadiationEnvironment}. The estimated energy fluxes are normalized to 1.0 at the trailing hemisphere apex.
(b) Calculated band area of the 2.07 \textmu m absorption feature for each of the 29 ``rings of constant radiolysis" considered. We find no measurable difference in the absorption depth between the chaos terrains and ridged plains. However, the absorption feature is essentially absent from Pwyll crater and its ejecta blanket.
(c) Calculated 2.07 \textmu m band area of the chaos terrains from panel b, with the chaos terrains further subdivided by type as described in \citet{leonard2023_EuropaGeologicMap}. There is no measurable difference in the 2.07 \textmu m band area between the different types of chaos terrain.
\label{fig:bandarea}}
\end{figure}

\section{Spatial Distribution}\label{sec:dist}
%\subsection{Tables} \label{subsec:tables}

With no clear resolution of the source of the 2.07 \textmu m feature from previous studies \citep{brown2013_SALTSRADIATIONPRODUCTS, ligier2016_VLTSINFONIOBSERVATIONS}, we turn to its spatial distribution in order to determine whether its presence is associated with the large-scale chaos terrains, which could confirm its suspected relationship to endogenic salts and possibly Europa's ocean chemistry. We note that endogenic salts are expected within Europa’s bands and ridges as well as within the chaos terrains \citep[e.g.][]{daltoniii2012_EuropaIcyBright}. Indeed, while the exact formation mechanism of these bands and ridges is not known, many proposed mechanisms require a direct link between the ice shell and ocean \citep[e.g.][]{head1999_EuropaMorphologicalCharacteristics, fagents2000_CryomagmaticMechanismsFormation,prockter2009morphology, johnson2017_PorositySaltContent, howell2018_BandFormationOceanSurface}. However, ridges on Europa are typically a few hundred meters to a few kilometers wide \citep{collins2009chaotic} and bands are no wider than 30 km \citep{prockter2002_MorphologyEuropanBands}. At the spatial resolution of the VLT ($\sim$130 km) the fractional coverage of a band or ridge feature within a single pixel is small. If irradiation of endogenic salts were responsible for the 2.07 \textmu m feature, we would therefore expect to measure a difference in the band strength between the large-scale chaos terrains which fully cover a single VLT spatial resolving element, and the background plains, which are overlaid by bands and ridges that only cover a small fraction of the spatial resolving element and are therefore unresolved. Indeed, the spectral signature of irradiated endogenic material correlated with the trailing hemisphere chaos terrains is seen in the maps of \citet{trumbo2020_EndogenicExogenicContributions} which have a comparable spatial resolution of $\sim$150 km. The 360 nm absorption feature attributed to radiolytically processed endogenic material, for example, is strongest within the large scale chaos terrains and is significantly weaker or absent from the non-chaos pixels on the trailing hemisphere. If confirmed, a similar association of the 2.07 \textmu m absorption band with Europa's large-scale chaos terrains would therefore support the hypotheses of \citet{brown2013_SALTSRADIATIONPRODUCTS} and \citet{ligier2016_VLTSINFONIOBSERVATIONS}, which suggest the absorption arises from some radiolytic product of endogenic material.

One common way to examine the spatial distribution of an absorption feature is to map its depth or area. However, the continuum shape between 2.0 and 2.2 \textmu m changes significantly across Europa's surface, largely due to variation in the strength of the 2.0 \textmu m water absorption band. As can be seen in Figure \ref{fig:specs}, the change in the slope, concavity, and overall spectral shape between the leading and trailing hemispheres is larger than the depth of the 2.07 \textmu m feature. The shape of the spectrum in this region varies gradually across Europa's surface, and we find that challenges in precisely defining this changing continuum, combined with significant pixel-to-pixel noise inherent in the IFU data, lead to large uncertainties in the measured band strength. Another option is to compare the depth of the 2.07 \textmu m feature between the chaos and non-chaos terrains integrated over the entire trailing hemisphere. However, the irradiation bullseye is symmetric about the trailing apex while the distribution of the chaos terrains on Europa's trailing hemisphere is not. As a result, the integrated chaos terrain spectrum samples areas on Europa with a sufficiently different continuum spectrum than is sampled by the integrated non-chaos pixels, such that it is difficult to determine whether these continuum differences wash out the signal of a geologic association or if the feature simply does not correlate with Europa's geology. We therefore construct a new technique wherein we consider rings of approximately constant radiolysis, where the overall continuum between $\sim$2 to 2.2 \textmu m is much more constant, and compare the strength of the 2.07 \textmu m feature for integrated spatial pixels that are within different large-scale geologic units and receive a similar irradiation dose.

To perform this spatial analysis, we first define the large-scale geologic units using the new United States Geologic Survey (USGS) global geologic map of Europa, which maps geologic units including craters, chaos terrains, bands, and regional plains at a scale of 1:15M \citep{leonard2023_EuropaGeologicMap}. Because the width of the bands and ridges are much smaller than our $\sim $130 km spatial resolution, we exclude these from our large-scale geologic units. We also exclude ten small craters with diameters less than $\sim$100 km and several small patches of discontinuous crater ejecta which are unresolved at the spatial resolution of the VLT. This leaves us with three distinct large-scale geologic units on Europa's trailing hemisphere - chaos terrains, ridged plains, and Pwyll crater including its large ejecta blanket. The chaos terrains can be further broken down into three sub-types - low albedo chaos, mottled chaos, and knobby chaos as defined in \citet{leonard2023_EuropaGeologicMap}. 

We then define a set of ``constant radiolysis rings" which cover the trailing hemisphere, where each ring contains pixels that are expected to receive a similar irradiation dose. While the idea is simple, the exact irradiation geometry on Europa is complex and remains an active area of research \citep{paranicas2009_EuropaRadiationEnvironment, cassidy2013_MagnetosphericIonSputtering, bagenal2020_SpaceEnvironmentIo}. The location dependent irradiation dose depends on a number of factors, including the type of particle, its energy, and the complex interaction between Jupiter's magnetosphere, the plasma torus, and Europa's ionosphere and a variety of irradiation patterns is possible \citep{bagenal2020_SpaceEnvironmentIo}. However, sulfuric acid production via the sulfur radiolytic cycle relies on the presence of sulfur ions from the plasma torus as well as radiolytic processing from energetic electrons and light ions, which are important sources of ionization energy that drive surface chemistry changes on Europa \citep{paranicas2001_ElectronBombardmentEuropa, paranicas2002_IonEnvironmentEuropaa, paranicas2009_EuropaRadiationEnvironment, cassidy2013_MagnetosphericIonSputtering}. The shape of the sulfuric acid bullseye is therefore a simple proxy for the total radiolytic processing from the combined effects of the plasma torus and magnetospheric particle bombardment occurring at a given location on Europa's trailing hemisphere.  To match the shape of the sulfuric acid bullseye, we start with a series of overlapping rings centered on the trailing hemisphere apex and defined by $\cos(\theta)$, where $\theta$ is the angular distance from the trailing hemisphere apex. We then empirically determine a latitudinal flattening coefficient of 0.6 so that the shape of the rings matches the sulfuric acid bullseye from Figure 10 in \citet{ligier2016_VLTSINFONIOBSERVATIONS} and set the width of the rings to match the diffraction-limited spatial resolution of the VLT. Figure \ref{fig:ring} shows a selection of four example rings distributed across the trailing hemisphere.

Once we have defined our constant radiolysis rings and large-scale geologic units, we assign each data pixel to its respective ring and geologic unit. Data pixels which overlap, typically because they were observed on two separate nights, are separated into multiple polygons and intersecting regions are averaged. In this way, we preserve the maximum possible spatial information from the dataset. In order to account for the wings of the point spread function (PSF), we exclude any spatial pixels whose central coordinates are within one spatial resolving element of a boundary between geologic units. Spatial pixels which cross the boundary of a ring are assigned to whichever ring intersects the largest fraction of the pixel. Pixels with more than 40\% of their area in two separate rings are duplicated and assigned to both. Once all of the pixels have been categorized, we sum the spectra from all of the pixels assigned to a given geologic unit and constant radiolysis ring. We then normalize these integrated spectra at 2.12 \textmu m, just past the absorption feature of interest, and plot the spectra for each geologic unit in order to compare the depth of the 2.07 \textmu m feature. Most of the rings only have integrated spectra for the chaos terrains and ridged plains, however a few rings also contain pixels from Pwyll crater and its ejecta blanket. The irradiation pattern is expected to be symmetric about the trailing hemisphere apex \citep{paranicas2009_EuropaRadiationEnvironment, hendrix2011_EuropaDiskresolvedUltraviolet, cassidy2013_MagnetosphericIonSputtering}, so we do not expect the east-west dichotomy in the terrain types, with significantly more chaos towards the sub-Jovian hemisphere and more plains towards the anti-Jovian hemisphere, to affect the results of our analysis. We also fit a fifth order polynomial continuum to each of the integrated spectra between 1.98 and 2.16 \textmu m, excluding the region between 2.04 and 2.1 \textmu m which contains the absorption feature and compare the continuum removed absorptions for the different geologic units. We choose a fifth order polynomial because it is able to provide a fair visual match to the continuum near the center of the trailing hemisphere as well as in the more icy regions where a third order polynomial produces an artificially deep absorption feature because the polynomial is not able to account for the concavity of the continuum shape, which is strongly influenced by the 2.0 \textmu m water ice absorption feature. We then continuum divide each spectrum and integrate the residual absorption between 2.05 and 2.1 \textmu m to compute an integrated band area.

As can be seen for a selection of four constant radiolysis rings in Figure \ref{fig:ring}, visual inspection of the integrated and continuum removed spectra (panels (b) and (c)) reveals that there is no discernible difference in the depth of the 2.07 \textmu m feature between the chaos terrains (red) and surrounding ridged plains (gray). Indeed, we find no discernible difference in the absorption band between the chaos and ridged plains in any of the 29 rings of constant radiolysis, which completely cover the trailing hemisphere. Likewise, we do not find a measurable difference in the integrated band areas between the chaos and ridged plains units for each ring, as seen in Figure \ref{fig:bandarea}. We do not include error bars on the measured band areas in Figure \ref{fig:bandarea} because we are unable to accurately calculate these uncertainties with the unknown effects of the changing continuum. However, we estimate a rough error of $\sim 10^{-4}$ on the integrated band area based on the scatter in the measured band areas of adjacent, overlapping rings. Figure \ref{fig:bandarea}(c) shows the measured band areas for each chaos sub-type, when present in each of the 29 constant radiolysis rings. Separately considering each chaos sub-type (low albedo chaos, mottled chaos, and knobby chaos) identified by \citet{leonard2023_EuropaGeologicMap} also reveals no discernible difference in the depth of the 2.07 \textmu m feature between the chaos terrain sub-types. Most of the trailing hemisphere chaos terrains are comprised of low albedo chaos ($\sim$64\% fractional coverage) and mottled chaos ($\sim$32\%), with only $\sim$4\% of the chaos at larger angular distances from trailing center classified as knobby chaos. 

It is interesting to note that the 2.07 \textmu m absorption feature is clearly absent from Pwyll crater and its ejecta blanket, even though Pwyll is located in a region with relatively high amounts of radiolysis. Ring 2 in Figure \ref{fig:ring} highlights the absence of this feature at Pwyll, and also illustrates that significant differences in the depth of the 2.07 \textmu m feature between different large-scale geologic units are identifiable with our spatial analysis technique. The absence of a 2.07 \textmu m absorption feature within the Pwyll crater ejecta blanket is also clearly seen in Figure \ref{fig:bandarea}(b), where the measured band areas for Pwyll are very small compared to the band areas of the chaos terrains and ridged plains within the same constant radiolysis rings.

While the strength of the 2.07 \textmu m absorption feature appears to be constant between the chaos terrains and ridged plains for each constant radiolysis ring, a clear decrease in the strength of the feature is seen as we move from the innermost rings with the highest expected radiolysis rates to the outermost rings where the 2.07 \textmu m feature is no longer seen. This decrease in band area is consistent with radiolytic production of the species responsible for the absorption feature. As a third and final check, we examined the division of the integrated ridged plains by the chaos terrain spectra and found that for some of the rings there was a broad dip in the ratioed spectrum between $\sim$ 2 to 2.2 \textmu m, which is a clear spectral match to water ice. This subtle difference in the continuum shape seen between the chaos terrains and ridged plains is consistent with a slight excess of water ice and therefore a deeper 2.0 \textmu m water ice absorption band in the ridged plains relative to the chaos terrains. We find no sign of any additional structure in the ratioed spectra between 2.05 and 2.1 \textmu m, further confirming that the 2.07 \textmu m absorption feature is unaffected by the compositional differences between the large-scale chaos terrains and ridged plains on Europa's trailing hemisphere.

\section{Discussion}\label{sec:discussion}

\subsection{Chaos Terrains and Ridged Plains}

As evidenced by its absence on Europa's leading, less-irradiated hemisphere, the species responsible for the 2.07 \textmu m feature is most likely a radiolytic product formed via sulfur radiolysis or processing via magnetospheric particle bombardment. Our spatial distribution analysis reveals that the depth of Europa's 2.07 \textmu m feature is spatially associated with the sulfuric acid bullseye on the trailing hemisphere, which we use as a proxy for the amount of radiolytic processing due to the combined effects of the plasma torus and magnetospheric particle bombardment. This distribution is consistent with a radiolytic origin for the absorption feature. However, we find no evidence of any spatial association with the large-scale chaos terrains which would be expected if the formation pathway required radiolytic processing of endogenic salty material. For any given ring of constant radiolysis, we find no measurable difference in the band area of the 2.07 \textmu m absorption feature between the large-scale chaos terrains and the icier ridged plains suggesting that the irradiation process which creates the 2.07 \textmu m absorption is independent of any endogenic material within Europa's recent geology. With both the epsomite hypothesis of \citet{brown2013_SALTSRADIATIONPRODUCTS} and the magnesium chlorates hypothesis of \citet{ligier2016_VLTSINFONIOBSERVATIONS} relying on the presence of irradiated endogenic magnesium chloride (\ch{MgCl2}), and no known exogenic source of magnesium ions to Europa \citep{bagenal2020_SpaceEnvironmentIo}, we find that the spatial distribution of the 2.07 \textmu m absorption band is inconsistent with both of these proposed origins for the absorption feature. Instead, our results suggest that the species responsible for Europa's 2.07 \textmu m absorption feature is a radiolytic product of water ice and exogenically sourced material and is not affected by the presence of whatever endogenic salt component exists within the chaos terrains.

\subsection{Pwyll Crater}

Pwyll crater has an $\sim$26 km diameter with a bright ray ejecta blanket that extends over 1000 km \citep{greeley1998_EuropaInitialGalileo}. Pwyll is thought to be young, with estimates for its age ranging from $\sim$ 3 - 18 Myr \citep{bierhaus2001_PwyllSecondariesOther}. The dark, red crater itself has apparently excavated endogenic material from a depth of $\sim$1 km \citep{garozzo2010_FateSbearingSpecies} and high spatial resolution Galileo/NIMS observations show evidence of asymmetric bands interpreted as salt-rich material at very high concentrations \citep{fanale2000_TyrePwyllGalileo}. However, the composition sharply transitions to be very ice-rich at the crater edge and throughout the ejecta blanket \citep{fanale2000_TyrePwyllGalileo}. At the spatial resolution of the VLT, we do not resolve the dark salty crater itself, and are dominated by the spectral signature of the bright ice-rich ejecta blanket. The distinct lack of a 2.07 \textmu m absorption feature within the Pwyll crater ejecta blanket, despite its presence in both the chaos terrains and ridged plains, implies that radiolytic processing has not yet had enough time to form the species responsible. The lack of a 2.07 \textmu m absorption within Pwyll may simply mean that the irradiation timescale for the process which produces the absorption feature is longer than the age of Pwyll crater. Or alternatively, there may not have been time for enough exogenic material from the plasma torus or magnetosphere to build up at Pwyll thus limiting the production of the species responsible for the absorption. In either case, the absence of the absorption within the Pwyll crater ejecta blanket places a strong lower limit of the age of Pwyll on the timescale over which the species responsible for the 2.07 \textmu m absorption forms on Europa. 

\subsection{Comparison with Irradiation Patterns}

Figure \ref{fig:bandarea}(b) shows the measured band area for each ring of constant radiolysis as a function of the angular distance due north from the trailing hemisphere apex. Panel (a) shows the expected shape of several distinct irradiation patterns on Europa's trailing hemisphere for comparison with the measured band areas. As previously noted, the absorption feature is absent from the Pwyll crater ejecta blanket, but both the chaos terrains and background ridged plains show a clear decrease in the band area with increasing angular distance. The measured fall-off in band area for the 2.07 \textmu m absorption feature with increasing distance from the trailing hemisphere apex is broadly consistent with a formation pathway driven by the irradiation pattern of the sulfur plasma, energetic ions and electrons, or the measured sulfuric acid bullseye. Cold sulfur plasma bombards the trailing hemisphere with a fall-off described by the cosine of the angle from trailing center \citep{hendrix2011_EuropaDiskresolvedUltraviolet, cassidy2013_MagnetosphericIonSputtering}, which we would expect to see reflected in the measured 2.07 \textmu m band depths if cold sulfur plasma controlled the formation of the absorption. If the formation pathway is instead driven by energetic electrons, we would expect a fall-off consistent with the shape of the power per unit area contours for 10 keV - 25 MeV electrons from Figure 8 in \citet{paranicas2009_EuropaRadiationEnvironment}. Or, if the formation pathway is driven by a combination of these irradiation patterns as expected for the sulfur radiolytic cycle, we might expect to find a fall-off consistent with the flattened cosine of the sulfuric acid bullseye. 

As can be seen in Figure \ref{fig:bandarea}(a), the differences in the shapes of these distributions are small near the center of the trailing hemisphere, where the 2.07 \textmu m absorption feature is strongest, and become more pronounced farther out where the feature has all but disappeared. Uncertainties in measuring the band depths for the various rings and a significant change in the continuum shape moving outwards from trailing center are sufficiently large so as to obscure the subtle differences which could differentiate between these irradiation signatures. The measured band areas for the chaos terrains and ridged plains in Figures \ref{fig:bandarea}(b) and (c) show a plausible match to all three irradiation patterns shown in panel a. Our results remain consistent with a variety of radiolytic production pathways and we are unable to determine whether the overall irradiation flux or the presence of any specific exogenic material is the limiting factor in the formation of the 2.07 \textmu m absorption.

\subsection{Potential Origins for the 2.07 \textmu m Absorption}

If the 2.07 \textmu m absorption feature does indeed arise from the radiolytic processing of water ice and exogenically sourced material on Europa's trailing hemisphere, the products of the known sulfur cycle offer an obvious first place to search for a plausible spectral match. Laboratory experiments have shown that various irradiation products can be created in electron or ion irradiated water ice bearing sulfur ions, or via sulfur ion bombardment of pure water ice. However, the specific products produced depend strongly on the temperatures, energies, and projectiles involved \citep{moore2007_RadiolysisSO2H2Sa, strazzulla2007_HydrateSulfuricAcidb, loeffler2011_RadiolysisSulfuricAcidb}. While hydrated sulfuric acid (\ch{H2SO4}) is expected to be the dominant irradiation product on Europa, some of the possible intermediate products include sulfur dioxide (\ch{SO2}), hydrogen sulfide (\ch{H2S}), and various sulfate anions (\ch{SO4^2-}, \ch{SO3^2-}, \ch{HSO4^-}) which can combine to form species such as sulfonic acid (\ch{SO2H2}) or hydrogen disulfide (\ch{H2S2}). \citet{tribbett2022_ThermalReactionsH2S} demonstrated that \ch{H2O + H2S + O3} ice mixtures at Europa temperatures can undergo thermal oxidation reactions on laboratory timescales, which may affect the steady state composition and intermediary products of the radiolytic sulfur cycle. 

Calculations based on laboratory experiments of sulfur ion implantation in water ice have shown that radiolysis can produce the expected concentration of sulfuric acid hydrate on Europa during $\sim 10^4$ years \citep{strazzulla2011_CosmicIonBombardment}. Therefore, if the feature arises from an intermediary of the sulfur cycle or a parallel irradiation process, it is somewhat surprising that we do not see any evidence for the 2.07 \textmu m absorption feature in Pwyll crater. However, Europa's UV to visible albedo ratio, which is anti-correlated with irradiation induced discoloration on the trailing hemisphere, also shows an enhancement at Pwyll suggesting that the discoloration timescale is longer than the age of Pwyll crater \citep{burnett2021_EuropaHemisphericColor} and consistent with the observed absence of the 2.07 \textmu m absorption feature. If the 2.07 \textmu m feature does arise from an intermediary of the radiolytic sulfur cycle, the lack of a feature at Pwyll could suggest that Pwyll crater is much younger than expected, that it has not yet had enough time to build up a sufficient amount of sulfur for sulfur radiolysis to occur, or alternatively that the calculations of \citet{strazzulla2011_CosmicIonBombardment} significantly underestimate the timescale over which the sulfur cycle should reach an equilibrium state. Future laboratory work may be crucial for understanding this discrepancy. 

Additionally, the 2.07 \textmu m absorption feature may also be explained by a parallel radiolytic cycle or some other unknown irradiation product. For example, the addition of \ch{CO2} ice into various sulfur radiolysis studies produced additional products such as carbonyl sulfide (\ch{OCS}), carbon disulfide (\ch{CS2}), carbonic acid (\ch{H2CO3}), and other carbon and sulfur bearing species \citep{garozzo2010_FateSbearingSpecies, mahjoub2017_ProductionSulfurAllotropes}. While \ch{CO2} has been detected on Europa \citep{mccord1998_NonwatericeConstituentsSurface, hansen2008_WidespreadCO2Other, carlson2009_EuropaSurfaceComp}, its spatial distribution across the trailing hemisphere is largely unconstrained. It is therefore uncertain whether an irradiation product of a combined carbon-sulfur radiolytic cycle is consistent with our observed spatial distribution, but nevertheless worth investigating.

We completed an extensive literature search of these potential sulfur- and carbon- bearing intermediary species and other possible irradiation products but were unable to find a plausible spectral match for the 2.07 \textmu m absorption amongst existing data sets. We are therefore unable to identify the source of the 2.07 \textmu m absorption feature, highlighting the need for additional irradiation experiments at Europa-like temperatures, particularly those including spectra across the full near-infrared wavelength range, in order to better understand the sulfur radiolytic cycle and determine whether any generated products can explain the 2.07 \textmu m absorption feature.

\section{Conclusions}

Using archived VLT/SINFONI H+K band spectra, we have shown that the presence and band area of the 2.07 \textmu m absorption feature on Europa's trailing hemisphere is not spatially associated with the large-scale geology, except for Pwyll crater and its ejecta blanket which lacks the absorption feature. There is, however, a spatial association between the 2.07 \textmu m absorption band and the trailing hemisphere sulfuric acid bullseye suggesting that the formation of the 2.07 \textmu m feature on Europa is independent of the endogenous salts thought to be present within the chaos terrains and is most likely an irradiation product of water ice and exogenic material. Thus, we find that neither epsomite nor a combination of magnesium chloride and magnesium perchlorate, as proposed by \citet{brown2013_SALTSRADIATIONPRODUCTS} and \citet{ligier2016_VLTSINFONIOBSERVATIONS}, respectively, are likely to explain the spatial distribution of the 2.07 \textmu m absorption feature and we consider an alternative hypothesis. We propose that the source of this feature may be an intermediary product of the radiolytic sulfur cycle, or something formed during the bombardment of water ice by electrons or the remaining (non-sulfur) ion species. Current laboratory data of relevant radiolytic species is fairly sparse in the 2.0 - 2.2 \textmu m range and we are unable to identify any plausible spectral matches. This highlights the need for more laboratory data at these wavelengths and at Europa-like conditions in order to identify radiolytic product(s) which may be responsible for the 2.07 \textmu m absorption feature and provide insights into the nature of the sulfur- and possibly carbon- radiolysis occurring on Europa.

%%%%%%%%%%%%%%%%%%%%%%%%%%%%%%%%%%%%%%%%%%%%%%%%%%%%%%%%%%%%%%%%%%%%%%%%%%%%%%
\vspace{12pt}
\begin{center}
\textbf{Acknowledgments}
\end{center}
This research has made use of the services of the ESO Science Archive Facility and is based on observations collected at the European Southern Observatory under ESO program 088.C-0833(A). M.R.D. would like to thank Dr. Erin Leonard for providing up-to-date shape files from the new United States Geologic Survey (USGS) global geologic map of Europa, which we used to define the various geologic units discussed in the text. S.K.T. is supported by the Heising-Simons Foundation through a 51 Pegasi b postdoctoral fellowship.
\newline


\vspace{5mm}
\facilities{VLT:Yepun (SINFONI)} % TO DO: check that this is the correct facilities macro
\software{Astropy \citep{astropy:2013, astropy:2018, astropy:2022},
          Cartopy \citep{Cartopy},
          GeoPandas \citep{geopandas},
          SciPy \citep{2020SciPy-NMeth},
          Shapely \citep{shapely2023},
          SpecUtils \citep{SpecUtils}
          }

%\appendix
%\section{Appendix information}
\bibliography{bibliography}{}
\bibliographystyle{aasjournal}

\end{document}